\documentclass[12pt,preprint]{aastex}

\newcommand{\msol}{\mathrm{M_{\odot}}}
\newcommand{\thco}{^{13}\mathrm{CO}}

\newcommand{\kms}{{\mathrm{km \, s^{-1}}}}
\newcommand{\de}{{^{\circ}}}
\newcommand{\ls}{\mathrel{\raise0.35ex\hbox{$\scriptstyle <$}\kern-0.6em
\lower0.40ex\hbox{{$\scriptstyle \sim$}}}}

\slugcomment{For submission to Ap J L }

\shorttitle{Scaleheight of GMCs}
\shortauthors{Stark and Lee}

\begin{document}

\title{The Scaleheight of Giant Molecular Clouds is\\
Less than that of Smaller Clouds}

\author{Antony A. Stark}
\affil{Smithsonian Astrophysical Observatory, Cambridge MA 02138}
\email{aas@cfa.harvard.edu}

\and

\author{Youngung Lee} 
\affil{Korea Astronomy Observatory, Taeduk Radio Astronomy
Observatory, Daejeon, Korea}
\email{yulee@trao.re.kr}

\begin{abstract}
We have used an antenna temperature thresholding algorithm 
on the Bell Laboratories $\thco$ Milky Way Survey to 
create a catalog of 1,400 molecular clouds. 
Of these, 281 clouds were selected for having
well-determined kinematic distances.  The scaleheight,
luminosity, internal velocity dispersion, and size of
the cloud sample are analyzed to show that clouds smaller than
$\sim  10^{5.5} \, \msol$ have a scaleheight which is
about 35 pc, roughly independent of cloud mass, while larger
clouds, the Giant Molecular Clouds, have a reduced
scaleheight which declines with increasing cloud mass.

\end{abstract}


\section{Introduction}

Since star formation occurs in molecular clouds,
the formation of molecular clouds is an essential first step 
to the star formation process.
The overall evolution of galactic metallicity
implies that hot ejecta from old stars must somehow
return to the cold molecular phase in order to
form new stars, and that the timescale for this process is
short compared to the age of the Galaxy \citep[e.g.][]{freeman02}.  
Giant Molecular Clouds (GMCs) are 
the largest concentrations of molecular
material, amounting to 
$2 \times 10^5 \msol$ or more in a region about
50 pc in size \citep{stark78}.
GMCs are strongly concentrated to spiral arms \citep{stark79c,lee01},
suggesting that they are transient phenomena:
hundreds of thousands of solar masses of molecular material 
collect during the passage of a spiral density wave,
and then dissipate some thirty million years later.
Thousands of stars are formed during the lifetime of the GMC, 
including the generators of the giant \ion{H}{2} regions
that trace spiral structure.
How molecular clouds form, and why they form, has been the subject of
ongoing theoretical investigation \citep{elmegreen00,pringle01}.
Computer simulations of cloud formation have become
increasingly complex and realistic
\citep{hartmann01,zhang02}.

The cloud formation process gathers and concentrates interstellar 
matter.  However this happens, we
expect that the random velocity of the resulting cloud
would be less than the velocity dispersion of the precursor
material, so that the velocity dispersion of GMCs as a class
would be less than that of smaller interstellar clouds.
This hypothesis is amenable to observational test.
The purpose of this {\em Letter} is to describe such
an observational test, and then to
quantify the effect in a way which may make for useful
comparisons with theory.

In \S 2, we take the Bell Laboratories $\thco$ Survey data and use a
brightness temperature thresholding algorithm to create a 
catalog of molecular clouds.
We then use our knowledge of Milky Way structure
in order to select a subset of clouds with well-determined
kinematic distances.  This allows us to determine the physical
size of the cloud and its distance above the galactic plane.
These data are analyzed in \S 3 to show that the scaleheight of
large clouds is less than that of smaller clouds.

\section{Cloud Identification and Selection}

The version of Bell Laboratories $\thco$ Survey  used
here is described by \citet{lee01}.  The survey covers
244 square degrees from 
$\ell = -5\de \, {\mathrm{to}} \, 117\de$, 
$b = -1\de \,{\mathrm{to}} \, +1\de$ 
sampled on a $3'$ or $6'$ grid with $\thco$ spectra having an
rms noise level of $T_R^{*} = 0.1\, {\mathrm K}$ in 
channels $0.68 \, \kms$ wide.  The survey data consists of
23 million data points (pixels) in a 3-dimensional $(\ell, \, b, \, v)$ space;
these are shown in \citet{lee01} as a series of $\ell$---$v$ maps.
The survey is not fully sampled, but it is sufficiently well-sampled
that no cloud within the survey volume larger than $\sim 1 \times 10^3 \msol$ 
would be missed due to undersampling.

Catalogs of clouds were generated from the survey data by 
a standard thresholding technique
\citep{solomon87,lee90,lee92}.
In this method, pixels having $T_R^{*} > T_{\mathrm{th}}$ are identified
for some value of a threshold brightness temperature $T_{\mathrm{th}}$.
The value of $T_{\mathrm{th}}$, typically 1, 2, or 3 K, is the
defining parameter of the catalog.  Pixels exceeding the
threshold are then grouped together in 
$(\ell, \, b, \, v)$ space to make a ``cloud", where all
the pixels constituting the cloud are above the
threshold and adjacent to at least one other pixel which
is also above the threshold.
What we mean by a ``cloud" is then a connected volume of pixels, all
of which are above the threshold.
This cloud recognition method has the advantage of
non-subjectivity---not only is it automated and unambiguous, it
makes no assumptions about cloud properties aside from the requirement
that a cloud be a connected region which is brighter than its surroundings.
A disadvantage of the thresholding method is that physically separated
molecular clouds can be bundled together into a single catalog
entry if they are projected onto each other 
such that above-threshold pixels within them appear to touch
from our point of view.  This happens more often for
small values of $T_{\mathrm{th}}$.
If the clouds are truly unassociated, the resulting
catalog entry will be systematically too large and have  
too high an internal velocity dispersion.
Another disadvantage is that all emission below $T_{\mathrm{th}}$ 
is discarded, cutting off the outer envelopes of the clouds and
systematically underestimating their total emissivity and extent.
This happens more often for large values of $T_{\mathrm{th}}$.
A choice of $T_{\mathrm{th}} = 1 \mathrm K $ for our $\thco$ data 
is an acceptable compromise.  This is approximately $10 \times$ the
rms noise level of the survey; since we require that a ``cloud" be
at least two pixels wide in each of the three dimensions
$(\ell, \, b, \, v)$, there are probably no cloud catalog entries referring to
non-existent clouds.
Fewer than 1\% of survey pixels exceed the 1 K threshold,
as discussed in \S2 of \citet{lee01},
so the number of above-threshold adjacencies occurring by 
chance will be small.
Applying the thresholding method with 
$T_{\mathrm{th}} = 1 \mathrm K $ on the
Bell Laboratories $\thco$ Survey
yields a catalog of 1,400 clouds.

We know the position of each cloud in $(\ell, \, b, \, v)$,
but we would also like to know its distance along the line
of sight in order to estimate its size and its height above
the galactic plane.  That can be done, to a greater or
lesser extent, by fitting the cloud's position
in $(\ell, \, b, \, v)$ to our knowledge of the velocity
field of the Galaxy.  Many molecular clouds are on a nearly
circular orbit around the Galactic Center at a velocity
which differs little from
$\Theta = 215 \, \kms$,
the rotation curve velocity.
Other clouds, particularly those near the Galactic Center,
can have significant non-circular motions.
For each cloud, there are various possible distances,
given our knowledge of motions in the Milky Way.
We can quantify this situation
by considering the set of possible distances.
Consider first the simple case of
a cloud in the second quadrant 
at $\ell = 110\de$, $b = 0\de$ and $v = -20 \, \kms$.
It is most likely to be at the point 2~kpc distant along the line of
sight corresponding to circular motion at that position and velocity,
since its total velocity probably does not deviate 
from the rotation velocity
by more than the typical cloud-cloud velocity dispersion. 
We can set approximate limits to range of possible distances
by first adding, then subtracting, the one-dimensional velocity
dispersion of $7 \, \kms$ \citep{stark89b} from its observed velocity,
finding that it is likely no more than 2.6 kpc distant nor
less than 1.3 kpc distant.  Next, consider a cloud in the first quadrant at, 
e.g., $l = 30\de$, which will have
two ranges of possible distances because of the rotation curve
distance ambiguity, resulting in four endpoints to the two (possibly
overlapping) ranges of distances.  
Another cloud with a large velocity which is 
projected onto the Galactic Center 
could be anywhere from 6 to 10 kpc distant.
Any cloud with a small velocity ($| v | < 14 \, \kms$) 
at any longitude could
be as near as 0.15 kpc or as distant as 1 kpc, or could be at other
distances permitted by the galactic rotation at its longitude and velocity.
Table 1 lists the ranges of possible distances for clouds in
various regions of $(\ell,  \, v)$ space.
Each cloud will fall within one or more of these regions,
resulting in a cumulative set of several possible distances.
Call the largest member of that set $d_{\mathrm{far}}$,
and the smallest member $d_{\mathrm{near}}$.
These values serve as approximate bounds on the distance to the
cloud, but they are not Gaussian errors---in most cases, the
true distance will be somewhere between these values, but the
distribution of true distances may be bimodal or asymmetric.
These limits do, however, allow us to select a
subset of the data whose distance ambiguity is not too large.

In the analysis below, we select only those clouds for which
$d_{\mathrm{far}}/d_{\mathrm{near}} < \sqrt{3}$.  This limit
is sufficiently large that it
does not exclude all clouds in the Galactic Center region.
Since luminosity varies as $d^2$, the ambiguity in the
luminosity of the distance-selected clouds is less than a factor of 3,
and so we know their luminosity within half an order of magnitude.
The distance-selected clouds are shown in color in
figure 1.  The selected sample consists of 281 clouds,
and they come from three general locations: 
the Perseus arm (at
$\ell > 80\de$ and $v < -14 \, \kms$), 
the first quadrant near the tangent velocity, and the Galactic Center.
Most of these clouds are several kiloparsec distant.
This is good, because one source of bias in the analysis
below comes from the limited range of the survey in galactic
latitude, $|b| < 1\de$.  We do not want this limit to exclude many clouds
that would otherwise be in the sample, and that will be true
only if most of the selected clouds are distant.  Figure 2 shows
that only a few of the distance-selected clouds are near the $b$
limit of the survey, while the unselected set of survey
clouds have a larger range and fill the survey space.

\section{Cloud Scaleheights}

The scaleheights of the distance-selected sample are shown in
figure 3, as a function of cloud luminosity.  Each cloud is
shown twice, once for the values corresponding to $d_{\mathrm{near}}$,
and once for the values corresponding to $d_{\mathrm{far}}$.
These points are connected by a line.  The true values for each
cloud probably lies somewhere near that line.  Here we have 
chosen to simply define scaleheight as the height above
the galactic equator at $b = 0\de$, rather than try to better
define the galactic midplane.  Also shown in this plot are
three very similar histograms, where the scaleheights 
have been averaged in bins an order of magnitude wide in
$\thco$ luminosity.  The dotted histogram only includes
values of $d_{\mathrm{near}}$, the dashed histogram only
includes values of 
$d_{\mathrm{far}}$, and the solid histogram includes both.
The essential similarity of these three histograms shows that
in the current sample of 281 clouds, the remaining distance
ambiguities are not important to the result.
For small, low luminosity clouds, the scaleheight is 
roughly constant at 35 pc, independent of cloud size.  
This value may be systematically small because of the
cutoff at $b = \pm1\de$, but it
is not significantly different from the scaleheight
of the overall CO brightness distribution
found by \citet{malhotra94a} using data with a much
larger range in $b$.
Note that
at a cloud luminosity of
about $10^4 \, \mathrm{K \, \kms \, pc^2}$, there is a break
in the distribution and the larger clouds have a reduced
scaleheight of about 20 pc for clouds in the range
$L(\thco) = 10^4$ 
to
$10^5 \, \mathrm{K \, \kms \, pc^2}$, and 5 or 10 pc
for the few clouds brighter than
$10^5 \, \mathrm{K \, \kms \, pc^2}$.
For the largest clouds, the derived scaleheight is 
smaller than the cloud's size and the galactic midplane
falls within the boundary of the cloud.

Since $\thco$ is in general not optically thick in these clouds,
the $\thco$ luminosity will be approximately proportional to
cloud mass \citep{lee94}.  This is 
illustrated in figure 4, where the luminosity
of the clouds in this sample is plotted against an estimate
of the mass obtained from the linewidth and the
cloud size.  
What we have done here is to ignore all 
terms in the magnetohydrodynamic virial equation
\citep{spitzer78} except for the kinetic and potential
energy volume integrals.  This is likely to be a valid
approximation, since the internal pressure in molecular
clouds is higher than the surrounding medium, 
the internal magnetic pressure is in approximate
equipartition with other pressures, and the timescale
for large-scale dissipation of the clouds is 
somewhat greater than a free-fall time. 
The absence of serious outliers in figure 4 is further evidence that 
the GMCs in our sample are real objects and not chance 
superpositions of small clouds.
The data are well-represented by the relation
$M_{\mathrm{vir}} = [20 \, \mathrm{\msol / K \, km \, s^{-1}}]
L(\thco)$, which corresponds to a 
galactic conversion factor \citep[e.g.,][]{sanders84}
$X(\thco) = 1.25 \times 10^{21} \mathrm{cm^{-2}/K \, km \, s^{-1}}$. 
The scaleheight breakpoint at a luminosity of
$10^4 \, \mathrm{K \, \kms \, pc^2}$ is therefore seen to correspond
to a cloud mass $ \sim 2 \times 10^5 \msol$,
the size of a small GMC.
The scaleheight of GMCs is less than that of
smaller clouds.  

\section{Conclusion} 

The data show that small molecular clouds have a scaleheight
which is approximately independent of cloud size.
Larger clouds, the GMCs, have a scaleheight which falls off
with mass.  This has been
demonstrated with a relatively small sample of clouds, but
it is a clean sample, chosen from a large-scale survey by
an algorithmic method.   
The implication is that the Giant Molecular Clouds, those
clouds which are concentrated in the spiral arms of the
Galaxy, are also concentrated to the galactic plane. 
We can understand this as a manifestation of the molecular cloud formation
process.
Atomic gas clouds, with a scaleheight of $\sim 100 \, \mathrm{pc}$
and a velocity dispersion $\sim 12 \, \kms$, condense out
of the diffuse atomic gas.
Their cores become molecular and increasingly more condensed,
resulting in a population of high-latitude ($\sim 70 \, \mathrm{pc}$), 
low mass ($M \ls 100 \, \msol$) partially-molecular clouds with
a dispersion $\sim 8 \, \kms$ \citep{malhotra94b,dame94}. 
These clouds become larger, more centrally condensed, and
more bound, with only a slight reduction in scaleheight 
and velocity dispersion, to
become molecular clouds ($ 100 \, \msol \ls M \ls 10^5 \, \msol$),
with dispersion $\sim 7 \, \kms$ \citep{stark89b}.  These
clouds can form stars, but are uniformly distributed throughout
the galactic disk, and may survive for many galactic rotations.
The largest clouds, the GMCs, are rapidly assembled 
by the passage of a spiral arm.  This process is
dissipative, different from the slow addition of material
that forms smaller molecular clouds.  
It results in a significant loss of random velocity
per unit mass, and the resulting GMCs are found at the galactic
midplane, in the spiral arm.  The ensuing formation of massive
stars destroys the cloud, fragmenting it into stars, ionized gas,
and small clouds
before the next interarm passage.

\acknowledgments

We thank the members of
the Bell Laboratories Radio Physics Research Group during the 
ten-year period of the Bell Labs $\thco$ Survey: R. W. Wilson, 
J. Bally, D. Mumma, W. Bent, W. Langer, G. R. Knapp,
and M. Pound.  
This work was supported by Basic Research Program
R01-2003-000-10513-0 of KOSEF, Republic of Korea,  
and 
by the William Rollins Endowment Fund 
of the Smithsonian Institution.

\bibliographystyle{apj}

\clearpage

\begin{deluxetable}{llcl}
\tabletypesize{\scriptsize}
\tablecaption{Possible cloud distances. \label{tbl-1}}
\tablewidth{0pt}
\tablehead{
\colhead{Location} & \colhead{$\ell$}   & \colhead{$v$}   & \colhead{Possible Distances} 
}
\startdata
all &\quad any &any &
$R_{\sun} \left( \mathrm{cos} \, \ell \, + \, \sqrt{\mathrm{sin^2} \, \ell \, [({v \over \Theta} + \mathrm{sin} \, \ell \,)^{-2} - 1]}\, \right)$\, ,\\
 & & &$R_{\sun} \left( \mathrm{cos} \, \ell \, + \, \sqrt{\mathrm{sin^2} \, \ell \, [({{v + v_\sigma} \over \Theta} + \mathrm{sin} \, \ell \,)^{-2} - 1]}\, \right)$\, ,\\
 & & &$R_{\sun} \left( \mathrm{cos} \, \ell \, + \, \sqrt{\mathrm{sin^2} \, \ell \, [({{v - v_\sigma} \over \Theta} + \mathrm{sin} \, \ell \,)^{-2} - 1]}\, \right)$\\
\\
$1^{\mathrm{st}}$ and $4^{\mathrm{th}}$ quandrants &$|\ell \,| < 90\de$ &
       $v  \, \mathrm{sin} \, \ell \, > 0$ &
$R_{\sun} \left( \mathrm{cos} \, \ell \, - \, \sqrt{\mathrm{sin^2} \, \ell \, [({v \over \Theta} + \mathrm{sin} \, \ell \,)^{-2} - 1]}\, \right)$\, ,\\
 & & &$R_{\sun} \left( \mathrm{cos} \, \ell \, - \, \sqrt{\mathrm{sin^2} \, \ell \, [({{v + v_\sigma} \over \Theta} + \mathrm{sin} \, \ell \,)^{-2} - 1]}\, \right)$\, ,\\
 & & &$R_{\sun} \left( \mathrm{cos} \, \ell \, - \, \sqrt{\mathrm{sin^2} \, \ell \, [({{v - v_\sigma} \over \Theta} + \mathrm{sin} \, \ell \,)^{-2} - 1]}\, \right)$\\
\\
Galactic Center region & $|\ell \, | < 8\de$ & any & ${5 \over 4}R_{\sun}  \, $,\\
 & & &${3 \over 4}R_{\sun}  $ \\
\\
3 kpc arm & $|\ell \, | < 11\de$ & $5 \ell -61\de < {v\over{1 \, \kms}} < 5 \ell -41\de$ & 
${5 \over 8}R_{\sun}  $\\
\\
$135 \, \kms$ arm & $|\ell \, | < 5\de$ & $5 \ell + 125\de < {v\over{1 \, \kms}} < 5 \ell + 145\de$ & 
${11 \over 8}R_{\sun} $ \\
\\
solar vicinity &\quad any &$|v| < 2 v_\sigma$ &0.15 kpc,\\
 & & &1 kpc \\
\enddata

\tablecomments{
The galactic parameters used are:
the Sun's distance to the Galactic Center,
$R_{\sun} = 8 \, \mathrm{kpc}$, 
the velocity of the flat rotation curve,
$\Theta = 215 \, \kms$, and
the one-dimensional velocity dispersion of molecular clouds,
$v_\sigma = 7 \, \kms$. 
Galactic longitude, $\ell$, expressed in degrees, 
ranges from $-180\de$ to $+180\de$.
Any possible distances which are not positive real numbers are discarded.
}

\end{deluxetable}

\clearpage

\begin{figure}
\plotone{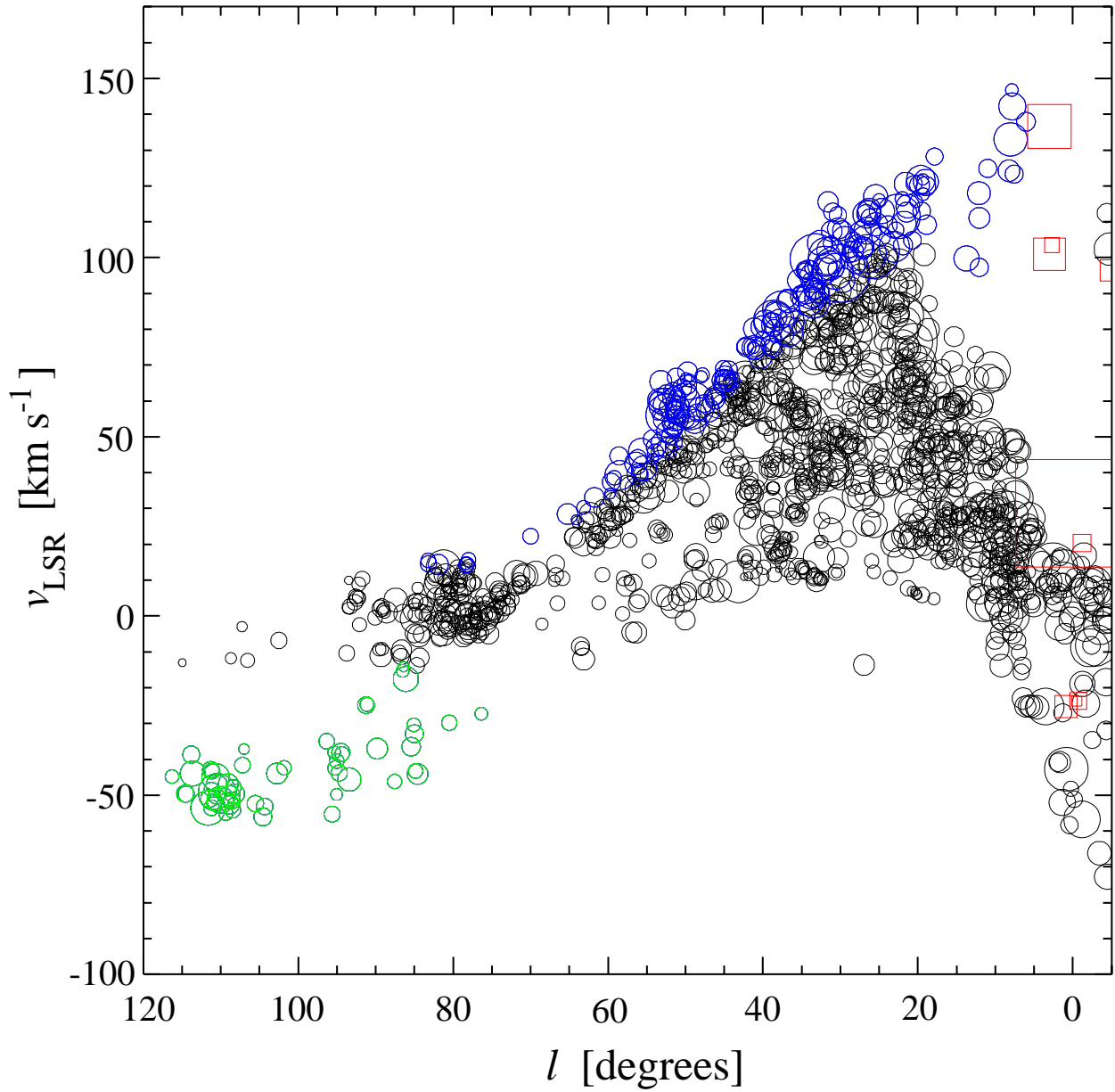}
\caption{Distribution in $\ell$ and $v$ of 1400 clouds 
from a cloud catalog
with $T_{th} = 1 \, \mathrm K$
made from the Bell Laboratories $\thco$ Survey.
The linear size of each symbol is proportional to the
velocity width of the corresponding cloud.
The 218 clouds which satisfy our distance accuracy requirement
are shown in color: red
squares for Galactic Center clouds, blue circles
for the molecular ring tangent velocities, and green 
circles for outer Galaxy clouds.
\label{fig1}}
\end{figure}

\clearpage 

\begin{figure}
\plotone{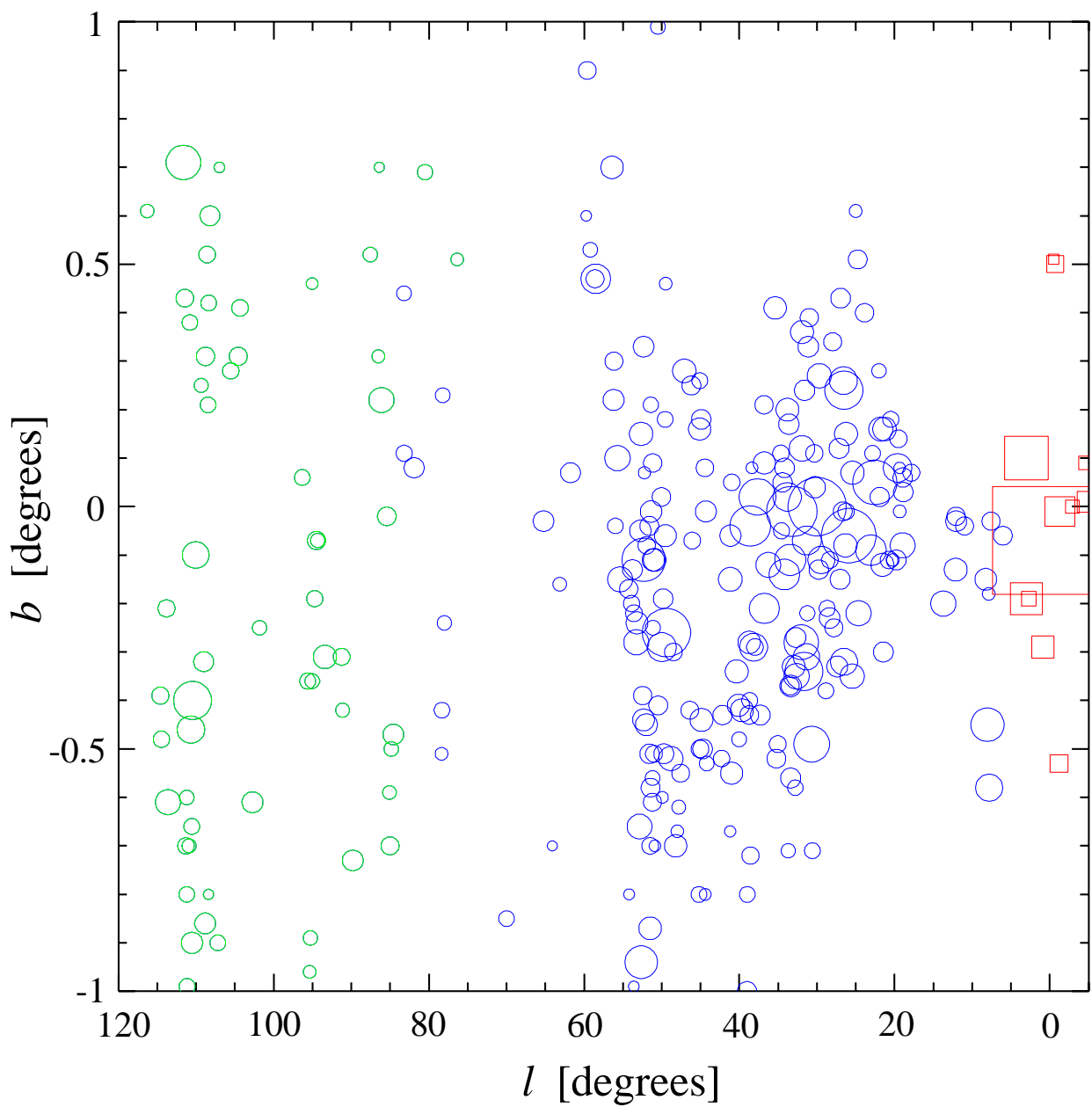}
\caption{Distribution in $\ell$ and $b$ of 281 distance-selected
clouds.
The clouds are color coded as in figure 1.
Note that the clouds as they appear on the sky are very much smaller
than the size of the symbols used here, especially in
the $\ell$ direction, and that the clouds are also
separated in velocity.  Chance superposition of
clouds is therefore unlikely.  The limit
of the survey in $b$ probably cuts off some clouds which should
be in the sample.
\label{fig2}}
\end{figure}

\clearpage 

\begin{figure}
\plotone{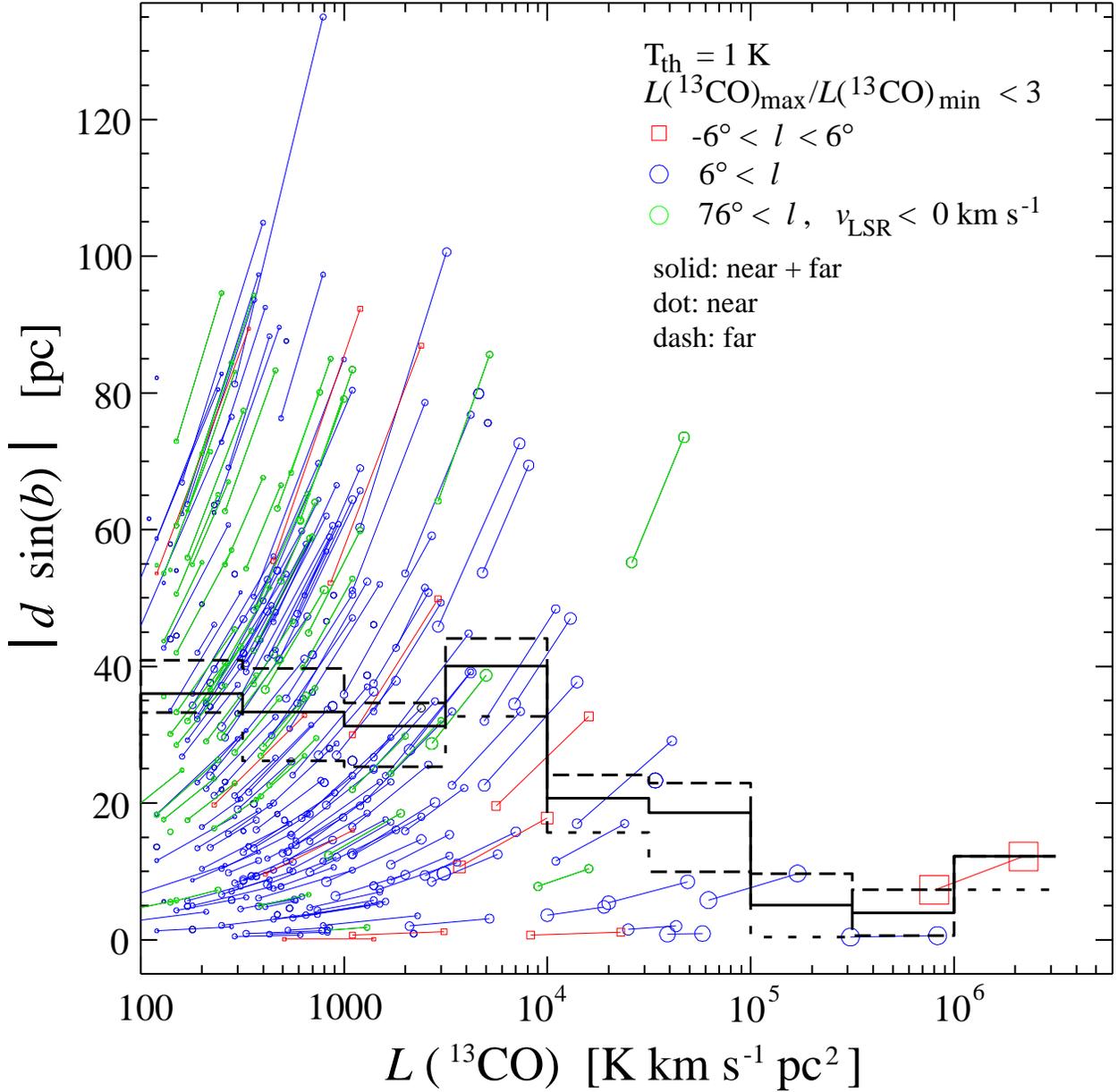}
\caption{Scaleheight vs. luminosity for 281 molecular clouds
selected for small distance uncertainty.
Each cloud is shown at its near distance and its far
distance, and these symbols are joined by a line.
The data points are averaged into bins half an order of magnitude
wide.  The averages are shown as three histograms: the dotted 
line includes only the near distance values, the dashed line 
includes only the far distance values, and the solid line
includes both.  The linear dimension of each symbol is proportional
to the cloud's linewidth.
\label{fig3}}
\end{figure}

\clearpage 

\begin{figure}
\plotone{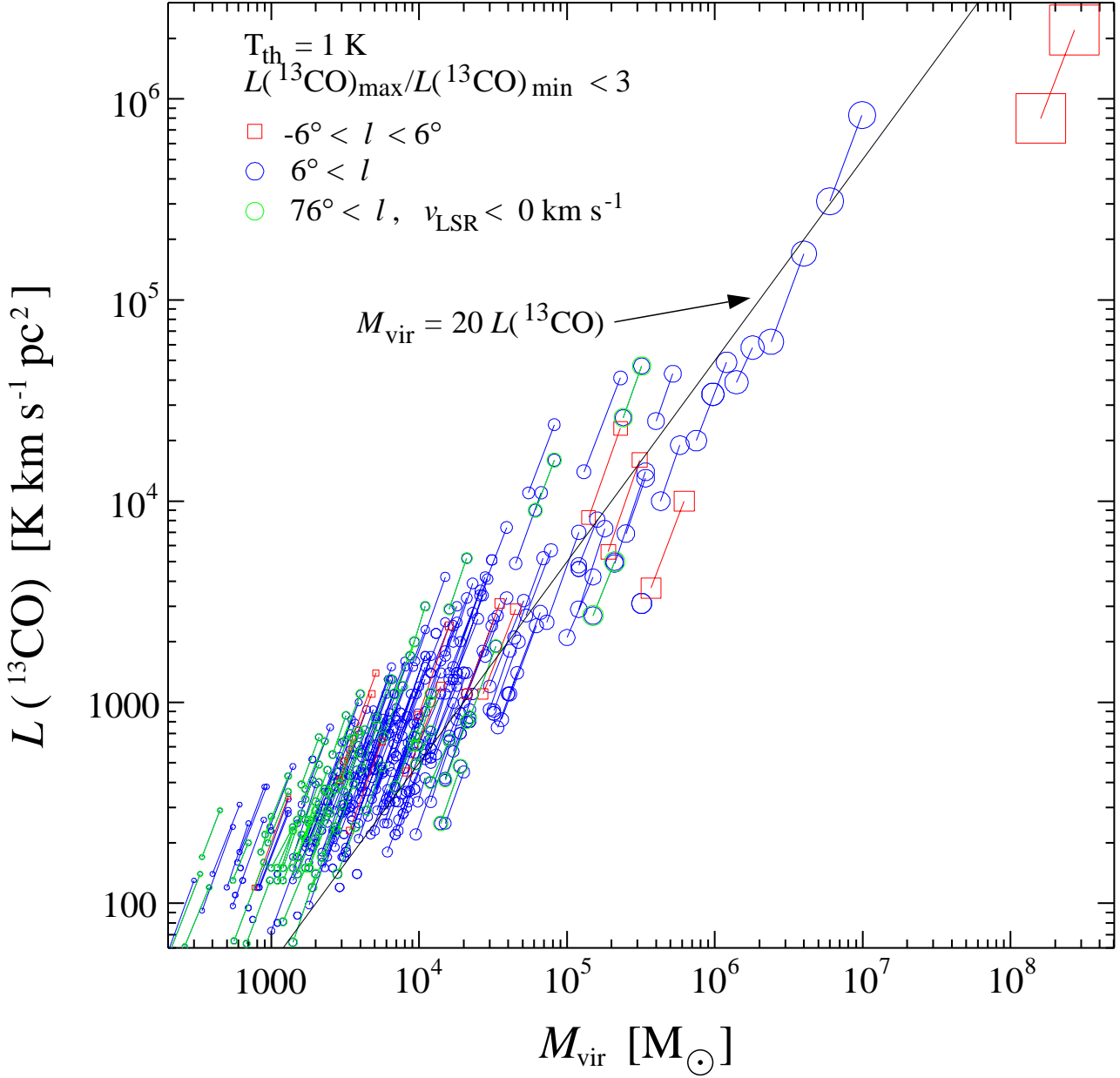}
\caption{Luminosity in $\thco$ vs. estimated virial 
mass for the 281 molecular clouds
selected for small distance uncertainty.
The linear dimension of each symbol is proportional
to the cloud's linewidth.  Galactic Center objects are shown as
squares.  The line $M_\mathrm{vir} = 20 \, L(\thco)$ is shown for
reference.
\label{fig4}}
\end{figure}

\clearpage

\end{document}